\documentclass[pra,aps,11pt]{revtex4-2}
\usepackage{graphicx,comment}
\usepackage{dcolumn}
\usepackage{bm,bbm}
\usepackage{soul}
\usepackage{yfonts}
\usepackage{amsmath}
\usepackage{float}
\usepackage{xcolor}

\def\F{{\boldsymbol{F}}}

\def\Id{{\mathbbm I}}
\def\Info{{\cal I}}
\def\BS{\rm BS}
\def\F{\mathcal{F}}
\usepackage{amsmath}
\usepackage{amssymb}
\usepackage{bm}
\usepackage[colorlinks=true, citecolor=blue, linkcolor=blue, urlcolor=blue]{hyperref}
\usepackage[a4paper,left=0.6in,right=0.6in,top=1in,bottom=1in]{geometry}
\usepackage{soul}

\begin{document}
\title{In situ characterization of a photon-subtraction device via heralding counts and homodyne detection}

\author{Priyanka Sharma}
\email{priyashara2016@gmail.com}
\thanks{Corresponding author}

\author{Matteo G.~A.~Paris}
\email{matteo.paris@fisica.unimi.it}

\author{Stefano Olivares}
\email{stefano.olivares@mi.infn.it}

\affiliation{Dipartimento di Fisica dell'Università di Milano, I-20133 Milan, Italy}

\begin{abstract}
    Photon subtraction is one of the most important techniques for generating non-Gaussian optical states and constitutes a key resource for quantum information processing and quantum metrology. The practical performance of a photon-subtraction device is primarily determined by the transmissivity of the beam splitter and the quantum efficiency of the heralding detector. Accurate knowledge of these parameters is therefore essential for assessing the quality of the generated non-classical states. In this work, we propose an experimentally feasible in situ scheme for the simultaneous estimation of these two parameters using only the measurement data produced during the operation of the device. Our protocol combines the click statistics of an on/off heralding detector with homodyne measurements performed on the transmitted mode of the beam splitter \textcolor{black}{when fed by a displaced squeezed state}. Within the framework of classical multi-parameter estimation theory, we derive the corresponding Fisher information matrix and investigate both joint and sequential estimation strategies. For simultaneous measurement of parameters, we evaluate the sloppiness of the underlying statistical model and analyze its dependence on the measured quadrature, probe photon number, squeezing fraction, beam splitter transmissivity, and detector efficiency. Our analysis proves that an appropriate choice of the homodyne quadrature substantially reduces parameter degeneracy and enables efficient simultaneous estimation. Furthermore, we show that the joint estimation strategy consistently provides a lower estimation bound than the sequential estimation approach over a broad range of experimentally relevant parameters.
\end{abstract}

\maketitle
\section{Introduction}
Non-Gaussian quantum states of light are a central resource for quantum information processing \cite{GenoniParis2010, barbieri2010nonGaussian,LACHMAN2022100395}. While Gaussian states \cite{parthasarathy2010gaussian}, such as squeezed vacuum \cite{teich1989squeezed, Schnabel2017} and coherent states \cite{PhysRev.131.2766}, are relatively easy to generate and manipulate, they are insufficient for achieving universal quantum computation \cite{NielsenChuang2010}, outperforming classical limits in quantum metrology, and testing fundamental quantum mechanics \cite{weedbrook2012gaussian, SHARMA2025130459}.  Among the most effective experimental techniques for producing such states from available Gaussian resources is photon subtraction \cite{Olivares2005PhotonSubtracted, Bellini2010, Zavatta2008}.
By removing a single photon from a squeezed vacuum state, one can herald the creation of a Schrödinger cat state \cite{Ourjoumtsev2007CatNature} which serves as a magic state for quantum error correction and fault-tolerant quantum computation \cite{ Guillaud2020, NielsenChuang2010}. More broadly, photon subtraction \cite{Zavatta2007, Zavatta2008} has been successfully employed to distill entanglement \cite{Takahashi2010}, enhance the fidelity of quantum teleportation, and generate non-classical correlations in hybrid quantum protocols \cite{Opatrny2000, NeergaardNielsen2011, NavarreteBenlloch2012}.

Optical implementation of photon-subtraction consists of a highly unbalanced beam splitter (transmissivity $\tau \simeq 0.9$--$0.99$) and a single-photon-sensitive on/off detector (e.g., an avalanche photodiode, APD) placed on the reflected beam \cite{javid2026comprehensive, pasharavesh2026photon, Bellini2010}. The input quantum state is sent into the main input port and a click on the on/off detector heralds the successful subtraction of a photon, with the output state available in the transmitted port \cite{Dakna1999,Ourjoumtsev2006,NeergaardNielsen2006, pasharavesh2026photon}. 
The practical performance of such a device is governed by two key parameters: the beam splitter transmissivity $\tau$, which determines the probability of subtracting a photon, and the quantum efficiency $\eta$ of the heralding detector, which directly affects the fidelity and the purity of the output state \cite{Dakna1999, Olivares2003}. The characterization of these parameters is crucial to assess the performance of the device and the amount of resources available for the subsequent quantum operations \cite{Paris2009,Giovannetti2011}.

The precise estimation of $\tau$ and $\eta$ is, however, nontrivial. Classical calibration methods, e.g., measuring $\tau$ with a strong coherent beam and $\eta$ with a calibrated power meter, can be performed independently, but they fail to capture the behavior of the device under low-photon-flux quantum conditions \cite{Giovannetti2011}. An \textit{in situ} characterization that relies solely on the data generated during actual quantum operation is therefore highly desirable \cite{Polino2020}.
In fact, the nominal values of the beam-splitter transmissivity and detector efficiency provided by manufacturers or determined via classical calibrations \cite{fbs23} are often inadequate for quantum applications \cite{PhysRevApplied.21.064059}. The transmissivity can exhibit slight wavelength or power dependencies, while the quantum efficiency of avalanche photodiodes is known to 
fluctuate with temperature, dark count rate, and the specific spatial mode of the incident field \cite{vir24}. Furthermore, classical calibration methods, which typically use bright laser beams, do not probe the device's behavior in the single-photon or few-photon regime, where nonlinearities or detector saturation effects can become relevant \cite{Akhlaghi:11}. This discrepancy between classical and quantum performance motivates the need for a characterization protocol that operates under the very same conditions as the intended quantum information task.
While full quantum state tomography \cite{d2003quantum,PhysRevX.7.031012} could, in principle, extract all relevant 
parameters of the output state, it usually requires a large set of measurements while offering limited precision \cite{addnoise97}, making it impractical for routine characterization.

In this work, we propose a feasible estimation scheme for the two parameters that uses only the data directly accessible from the device itself: (i) the click statistics of the on/off heralding detector, and (ii) homodyne detection measurements performed on the subtracted (transmitted) beam. The latter provides access to the quadrature distribution of the output state conditioned on a heralding click. Our scheme does not aim for ultimate precision allowed by quantum mechanics or the derivation of novel Cramér–Rao bound. Rather, it aims to provide an experimentally accessible, method to extract $\tau$ and $\eta$ from a realistic dataset \cite{Sharma_2025}. By combining the information from the detector counts (which constrain the product $\eta\times\tau$) and the homodyne distribution of the output state (which reveals the conditional statistics of the subtracted field), one is indeed able to decouple the two parameters unambiguously.
Our proposed scheme is specifically designed to be minimally invasive, using the very data  that would be generated in a standard photon-subtraction experiment. This allows for a quick, on-demand calibration with no extra experimental overhead, facilitating the use of such devices in larger quantum photonic circuits.

The paper is structured as follows. In Section  \ref{Methodology}, we establish the theoretical framework, deriving the conditional Gaussian state of the transmitted mode and the corresponding Fisher information matrix that combines information from both the detector's click statistics and the homodyne measurements. This allows us to introduce the concept of sloppiness as figure of merit for quantifying parameter degeneracy \cite{10.1063/1.4923066}. In Section \ref{Discussion}, we present our results, first identifying the optimal homodyne quadrature for estimation and then analyzing how the probe-state resources (photon number and squeezing fraction) influence the sloppiness.  We also compare joint estimation strategies of both the parameters to sequential measurement, and discuss the results: Our findings consistently show that the joint strategy offers a superior estimation bound across a broad range of experimentally relevant parameters. Section \ref{Conclusion} closes the paper with some concluding remarks. 

\section{Theoretical Framework}\label{Methodology}
\begin{figure}
\centering
\includegraphics[width=0.35\textwidth]{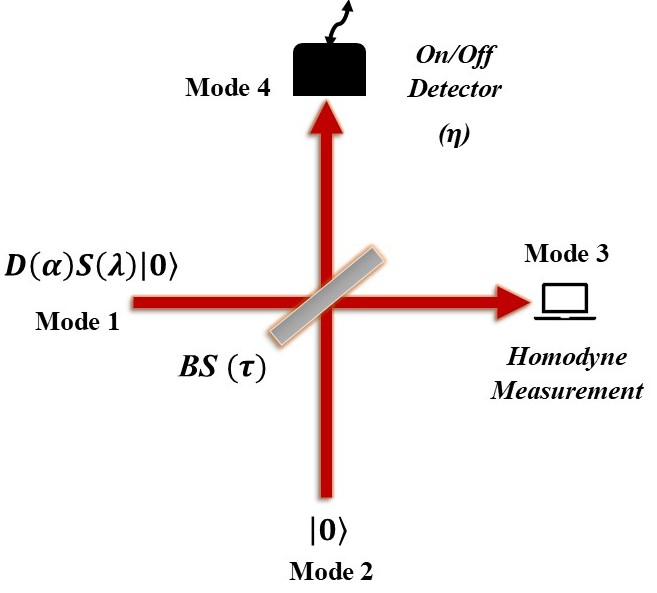}
\caption{\label{fig_1}
Sketch of the estimation setup. A displaced squeezed state (mode 1) is mixed with the vacuum state (mode 2) at a beam splitter (BS) with transmissivity $\tau$. Homodyne detection is performed on the transmitted mode 3 while the reflected mode 4 is measured by an on/off detector with quantum efficiency $\eta$. The goal is to jointly estimate $\tau$ and $\eta$ from the detector's statistics.
}
\end{figure}

We consider a single-mode \textcolor{black}{optical field} prepared in a displaced squeezed state
$|\alpha,\lambda\rangle = \hat{D}(\alpha)\hat{S}(\lambda)|0\rangle$, \textcolor{black}{where $\alpha \in {\mathbbm C}$ and $\lambda\in \in {\mathbbm R}$ are the displacement amplitude and the squeezing parameter, respectively. The probe state (mode 1) is mixed with the vacuum (mode 2) at a beam splitter (BS) with unknown transmissivity $\tau$, as illustrated in Fig.~\ref{fig_1}.} One of the output modes is monitored by an on/off detector (mode 4) characterized by a quantum efficiency $\eta$. On the other output (mode 3) we  perform homodyne detection \cite{sparaciari2015bounds}. 

To estimate these parameters, we employ the framework of (classical) multi-parameter estimation theory \cite{Yang:19,Qiongyi2026DistributedQuantumSensing, PhysRevA.105.022406}. For a parameter vector
$\boldsymbol{\theta} = (\tau,\eta)$, the precision of any unbiased estimator is bounded by the classical Cramér-Rao inequality \cite{Sharma_2025, Holevo, Paris2009,PhysRevA.78.032303}
\begin{equation}
    \mathrm{Cov}(\boldsymbol{\theta}) \geq \frac{\mathcal{F}^{-1}}{M},
    \label{classicalCRB}
\end{equation}
where $\mathrm{Cov}(\boldsymbol{\theta})$ is the covariance matrix of the estimators, $\mathcal{F}$ is the classical Fisher information matrix, and $M$ denotes the number of measurements.

In general, the elements of the Fisher information matrix are given by \cite{Paris2009, ALBARELLI2020126311}
\begin{align}
 \left[ \mathcal{F}\right]_{jk}
    & =
    \sum_{\chi}
    p(\chi|\boldsymbol{\theta})
    \big[\partial_j \ln p(\chi|\boldsymbol{\theta})\big]
    \big[\partial_k \ln p(\chi|\boldsymbol{\theta})\big]\nonumber \\
    & =\mathbb{E}\Big[\big[\partial_j \ln p(\chi|\boldsymbol{\theta})\big]
    \big[\partial_k \ln p(\chi|\boldsymbol{\theta})\big]\Big]
    \label{Fij}
\end{align}
where $p(\chi|\boldsymbol{\theta})$ is the probability of obtaining measurement outcome $\chi$, and $\partial_j \equiv \partial/\partial \theta_j$. {Extension to a continuous outcome is straightforward.}

A convenient scalar figure of merit for joint estimation is the total variance bound
\begin{align}
\hbox{Tr}[\mathrm{Cov}(\boldsymbol{\theta})] & \geq \frac{C_F}{M} \\
    C_F & = \mathrm{Tr}\!\left[\mathcal{F}^{-1}\right],
\end{align}
which quantifies the minimum achievable sum of variances of both estimated parameters. A smaller value of $C_F$ corresponds to better overall estimation precision. It should however be noticed that in 
multi-parameter models, the information about different parameters may be distributed very unevenly. This phenomenon may lead to the phenomenon of \emph{sloppiness} \cite{10.1063/1.4923066,ALBARELLI2020126311,liu2020quantum,frigerio2024overcomingsloppinessenhancedmetrology, PhysRevLett.121.130503, PhysRevA.94.052108, Candeloro_2024}. Sloppy models are characterized by a Fisher information matrix with a vanishing determinant or strongly unequal eigenvalues, indicating that some combinations of parameters are poorly identifiable. To quantify this effect, we employ the sloppiness parameter 
\begin{equation}
S = \frac{1}{\det (\mathcal{F})}.\label{Sloppi}
\end{equation}
A large value of $S$ signals a nearly singular Fisher information matrix and hence strong parameter degeneracy, whereas a small value of $S$ indicates that both parameters can be estimated simultaneously.

In our scheme, the estimation protocol relies on the statistics of the transmitted field conditioned on the heralding outcome, and therefore the first step is to determine the conditional Gaussian state after the beam splitter. In particular, we are going to derive the covariance matrix and displacement vector of the transmitted mode following the Gaussian state and operation formalism.

\subsection{Evolution and Conditional Measurement}
\label{Sec:Evolution}
The covariance matrix of the initial squeezed state and the corresponding displacement vector are:
\begin{equation}
\gamma_1 = \frac{1}{2}
\begin{pmatrix}
e^{-2\lambda} & 0 \\
0 & e^{2\lambda}
\end{pmatrix},
\qquad
d_1 =
\begin{pmatrix}
x_0 \\
y_0
\end{pmatrix},
\end{equation}
with,
\begin{align}\label{x0:y0}
x_0 = \sqrt{2}\alpha_x, \qquad
y_0 = \sqrt{2}\alpha_y,
\end{align}
and we put $\alpha = \alpha_x + i\alpha_y$. 
In turn, the covariance matrix and the displacement vector of the vacuum states read:
\begin{equation}
\gamma_2 = \frac{1}{2}\,\Id_2,
\qquad
d_2 =
\begin{pmatrix}
0 \\
0
\end{pmatrix}.
\end{equation}
where $\Id_2$ is the $2\times2$ identity matrix. 
The initial displaced squeezed state is mixed with the vacuum in a BS, described by the unitary
transformation $U_{\BS}$. Since the mode transformation is linear \cite{olirev2012}, the output state remains Gaussian, and the corresponding covariance matrix and displacement vector are obtained by applying the symplectic transformation of Eq.~(\ref{BS}). The detailed derivation is presented in Appendix~\ref{AppA}.
Photon subtraction is heralded by an on/off detector placed in the reflected arm. The positive operator-valued measure (POVM) element associated with the ``off'' outcome is \cite{olirev2021}
\begin{equation}
\Pi_0
=
\sum_{k=0}^{\infty}
(1-\eta)^k
|k\rangle\langle k|,
\label{POVM}
\end{equation}
which corresponds to a Gaussian measurement. The on/off detector does not click with probability
\begin{align}
P(0|\mathbf{\theta}) =   \hbox{Tr} \left[ U_{\BS}
|\alpha,\lambda\rangle\langle\lambda,\alpha| U_{\BS}^\dag\, \Id\otimes\Pi_0
\right]
\end{align}
and the corresponding ``off" conditional state 
of the transmitted mode is given by
\begin{align}
\rho_0  = \frac{1}{P(0|\mathbf{\theta})}\hbox{Tr}_2 \left[ U_{\BS}
|\alpha,\lambda\rangle\langle\lambda,\alpha| U_{\BS}^\dag\, \Id\otimes\Pi_0
\right]\,,
\label{cond0}
\end{align}
which is still a Gaussian state with covariance matrix 
(see Appendix~\ref{AppA} for details about the derivation)
\begin{equation}
\gamma_0
=
\frac12
\begin{pmatrix}
\Gamma_q&0\\
0&\Gamma_p
\end{pmatrix},
\label{condcov}
\end{equation}
where
%
\label{Gamma}
\begin{align}
\Gamma_q
= 1 + \gamma_{-}
\quad\mbox{and}\quad
\Gamma_p = 1 + \gamma_{+},
\end{align}
and displacement vector
\begin{equation}
d_0
=
\begin{pmatrix}
\langle q \rangle \\[1ex]
\langle p \rangle 
\end{pmatrix} = 
\sqrt{\tau} \begin{pmatrix}
\, x_0 \left(1-\gamma_{-}\right) \\[1ex]
\, y_0 \left(1-\gamma_{+}\right) 
\end{pmatrix},
\label{conddisp}
\end{equation}
where $x_0$ and $y_0$ are given in  Eq.~(\ref{x0:y0}), and
\begin{equation}
\gamma_{\pm} = \frac{\eta  \left(e^{\pm 2 \lambda }-1\right) (1-\tau)}{2+\eta(1-\tau)(e^{\pm 2\lambda}-1)}.
\end{equation}

Now that the conditional state has been obtained, the next step is to determine how much information about the unknown parameters can be extracted from homodyne data. This can be quantified by evaluating the
associated Fisher information matrix conditioned on the ``off" event.

\subsection{Fisher Information of the homodyne measurements for the ``OFF'' Events}
\label{Sec:FIOFF}
Conditioned on the ``off'' outcome of the heralding detector, the transmitted mode remains Gaussian. Consequently, the quadrature probability distribution obtained by homodyne detection is also Gaussian, 
allowing us to evaluate the Fisher information analytically.
Given a Gaussian probability distribution with parameter-dependent mean $\mu_\phi$ and variance $\sigma_\phi^2$, the corresponding Fisher information matrix is
\begin{equation}
\Info^{{\rm hd(off)}}
=
\begin{pmatrix}
\Info_{\tau\tau} & \Info_{\tau\eta}\\[1ex]
\Info_{\eta\tau} & \Info_{\eta\eta}
\end{pmatrix},
\label{FHom}
\end{equation}
with matrix elements
\begin{equation}
\left[\Info^{{\rm hd(off)}}\right]_{jk}
= \frac{\partial_j\mu_\phi\,
\partial_k\mu_\phi}{\sigma_\phi^2}
+
\frac{\partial_j\sigma_\phi^2\,
\partial_k\sigma_\phi^2}{2\sigma_\phi^4},
\label{FGaussian}
\end{equation}
and, using the conditional moments obtained in Eqs.~(\ref{condcov})--(\ref{conddisp}), we find (see Appendix~\ref{AppC1} for further detail about the calculation):
\begin{align}
\mu_\phi
&=
\langle p_\phi\rangle =
\langle q \rangle \cos{\phi} + \langle p \rangle \sin{\phi}
\label{MeanQuadrature}
\\[1ex]
\sigma_\phi^2
&=
\frac12
\left(
\Gamma_q\cos^2\phi
+
\Gamma_p\sin^2\phi
\right).
\label{VarianceQuadrature}
\end{align}

The analytical treatment above applies only to the ``off" events, where the conditional state is Gaussian. Now we perform the calculation for the complementary ``on" events. However, photon subtraction produces a non-Gaussian conditional state, requiring a different evaluation of the Fisher information matrix.

\subsection{Fisher Information of the homodyne measurements for the ``ON'' Events}
\label{Sec:FION}

\textcolor{black}{The POVM element corresponding to the heralding click follows from Eq.~(\ref{POVM}) and writes}
\begin{equation}
\Pi_1
= \mathbb{I}-\Pi_0\,,
\end{equation}
such that $P(1|\mathbf{\theta}) = 1 - P(0|\mathbf{\theta})$.
The corresponding conditional output state associated with the ``on'' ($k=1$) event is
\begin{equation}
\rho_1 =
\frac{\bar\rho-P(0|\boldsymbol{\theta})\rho_0}
{1-P(0|\boldsymbol{\theta})},
\label{rhoON}
\end{equation}
where 
\begin{align}
\bar\rho = 
\hbox{Tr}_2 \left[ U_{\BS}
|\alpha,\lambda\rangle\langle\lambda,\alpha| U_{\BS}^\dag\, \Id\otimes\Id
\right]\,,
\end{align} denotes the unconditional transmitted state (i.e. the state of mode $3$ if we ignore the result of the on/off detection) and $\rho_0$ is the conditional Gaussian state associated with the
``off'' outcome ($k=0$), i.e., the state in Eq. (\ref{cond0}). 
The
corresponding conditional homodyne distribution becomes
\begin{equation}
p_1(x|\boldsymbol{\theta})
=
\frac{
\bar{p}(x|\boldsymbol{\theta})-P(0|\boldsymbol{\theta})p_0(x|\boldsymbol{\theta})
}
{1-P(0|\boldsymbol{\theta})},
\label{ONdistribution}
\end{equation}
with
$\bar{p}(x|\boldsymbol{\theta})
=
{\rm Tr}
\left[
\bar{\rho}\,
|x_\phi\rangle
\langle x_\phi|
\right]$ and 
$p_0(x|\boldsymbol{\theta})
=
{\rm Tr}
\left[
\rho_0\,
|x_\phi\rangle
\langle x_\phi|
\right]$. Remind that both $\bar{p}(x|\boldsymbol{\theta})$ and  $p_0(x|\boldsymbol{\theta})$ are Gaussian distributions.


The probability (\ref{ONdistribution}) is a non-Gaussian distribution since it is the normalized difference of two Gaussian probability distributions. Consequently, unlike the ``off'' case, no closed explicit expression
exists for the Fisher information. The Fisher information associated with the ``on'' events is therefore evaluated directly from the definition given in Eq.~(\ref{Fij}) using Eq.~(\ref{ONdistribution}). Eventually, the Fisher information matrix elements are given by 
\begin{align}
   \left[\Info^{{\rm hd(on)}}\right]_{jk}
    & =
    \int dx\,
    p_1(x|\boldsymbol{\theta})
    \big[\partial_j \ln p_1(x|\boldsymbol{\theta})\big]
    \big[\partial_k \ln p_1(x|\boldsymbol{\theta})\big] .
\end{align}

The complete estimation protocol must combine the information extracted from both conditional branches together with the information contained in the detector click statistics itself, leading to the total Fisher information matrix.

\subsection{Total Fisher Information}
\label{TotalFI}
Given the parameter vector $\boldsymbol{\theta}=(\tau,\eta)$,
the overall measurement record consists of the dichotomous detector outcome $k\in\{0,1\}\equiv\{\mbox{``off''},\mbox{``on''}\}$, with probability $P(k|\boldsymbol{\theta})$, together with the continuous homodyne detection outcome $x\in {\mathbbm R}$, with probability $p_k(x|\boldsymbol{\theta})$ conditioned to on/off event $k$. The total Fisher information matrix elements thus reads (see Appendix~\ref{AppC2} for details about the derivation)
\begin{align}
\left[\mathcal{F}\right]_{jk}
&= \sum_{\chi=0,1} P(\chi|\boldsymbol{\theta})\, \left[ \Info^{{\rm hd}(\chi)}\right]_{jk}
+ \left[\mathcal{H}^{\rm on/off}\right]_{jk},
\label{TotalFisher}
\end{align}
where $\left[\Info^{{\rm hd}(\chi)}\right]_{jk}$ denotes the Fisher information matrix elements associated with the homodyne measurement conditioned on outcome $\chi$, while
$\left[\mathcal{H}^{\rm on/off}\right]_{jk}$ represents the information contained in the detector click statistics.

\section{Results and Discussion}\label{Discussion}
In this section, we investigate the simultaneous estimation of the beam-splitter transmissivity $\tau$ and the quantum efficiency of the heralding detector $\eta$ using the total Fisher information matrix $\left[\mathcal{F}\right]_{jk}$ derived in Sec.~\ref{TotalFI}. The total mean photon number of the probe state is given by
\begin{equation}
N=|\alpha|^2+\sinh^2\lambda,
\end{equation}
where $|\alpha|^2$ denotes the coherent contribution and $\sinh^2\lambda$ represents the photon number generated by squeezing. To conveniently distribute the available resources between these two components, we introduce the dimensionless squeezing fraction
\begin{equation}
\beta = \frac{\sinh^2\lambda}{N}.
\end{equation}

 \begin{figure}[h]
\centering
\includegraphics[width=.33\columnwidth]{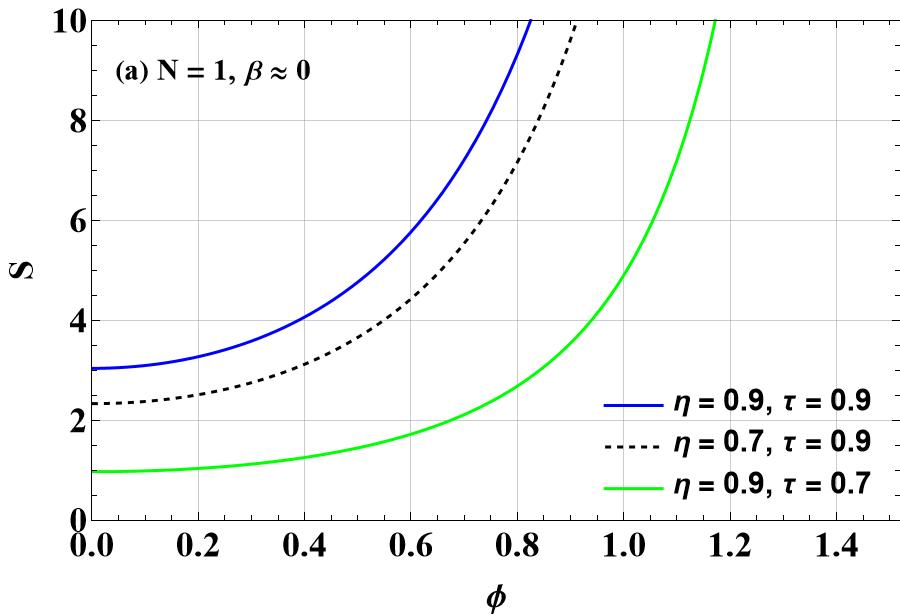}
\includegraphics[width=.33\columnwidth]{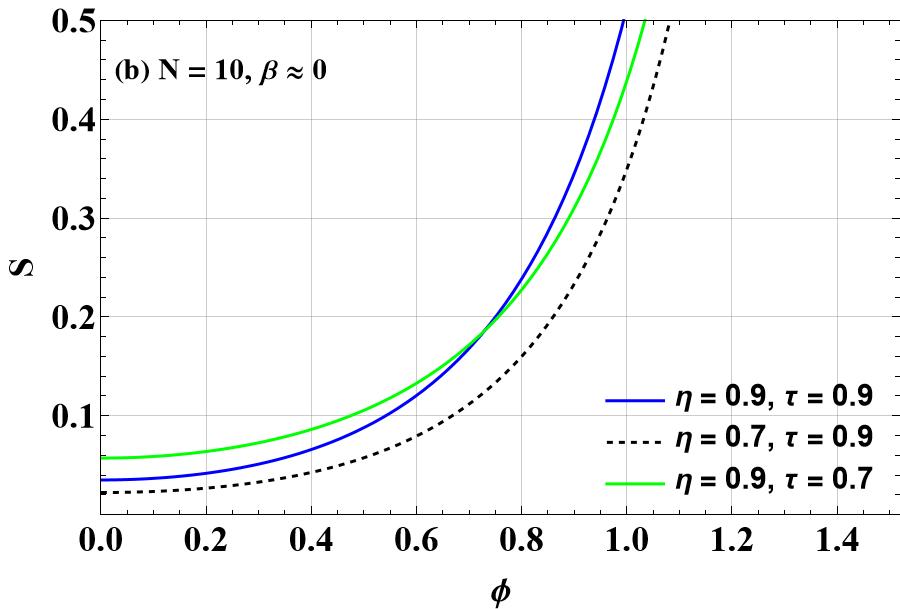}
\includegraphics[width=.33\columnwidth]{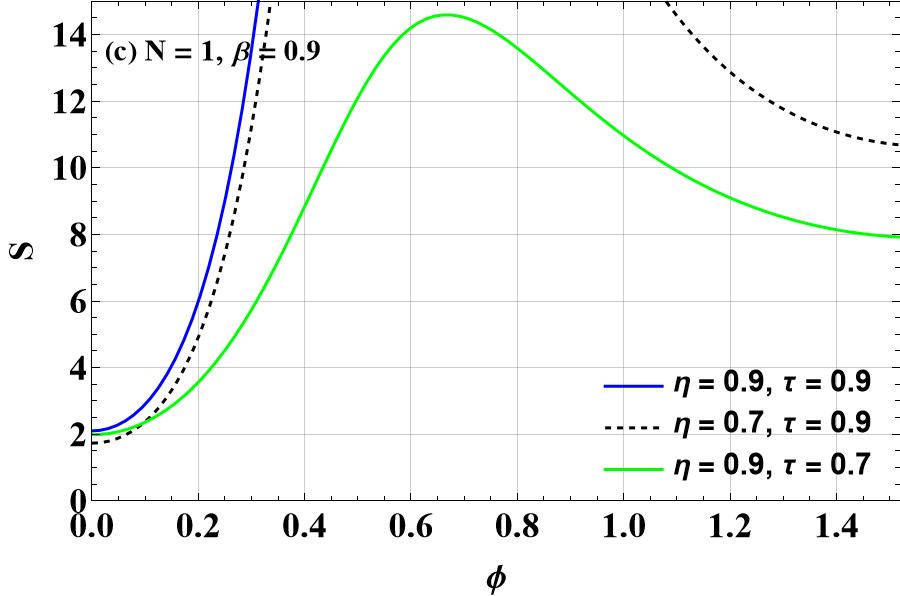}
\includegraphics[width=.33\columnwidth]{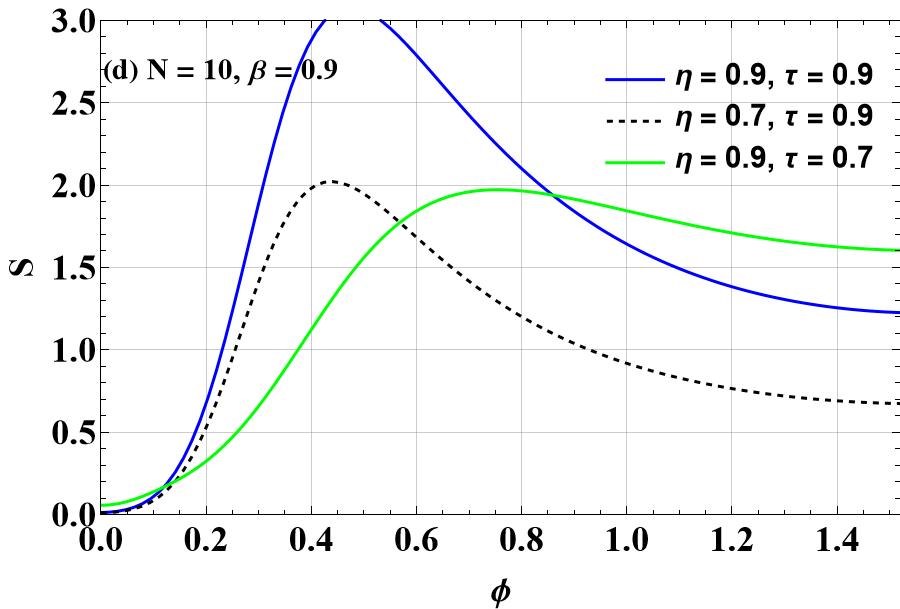}
\caption{\label{fig_2} Sloppiness $S$ as a function of the homodyne phase $\phi\in [0,\pi/2]$, for different values of $\tau$ and $\eta$. Panels (a) and (b) show the results in the absence of squeezing ($\beta\approx0$) with $N=1$ and $N=10$, respectively, whereas panels (c) and (d) correspond to a squeezing fraction $\beta=0.9$ with $N=1$ and $N=10$, respectively. In all the plots we can clearly identify the minimum at $\phi=0$, corresponding to the amplitude quadrature measurement.}
\end{figure}
A central quantity in our analysis is the sloppiness parameter $S$, defined in Eq.~(\ref{Sloppi}), characterizing the identifiability of the estimated parameters by quantifying how uniformly the Fisher information is distributed among different parameter directions. A smaller value of $S$ indicates that the information is more evenly distributed, resulting in improved simultaneous estimation.

Firstly, we determine the optimal measurement configuration by examining the dependence of the sloppiness parameter on the homodyne phase $\phi$ (due to the periodicity of the homodyne outcomes, we just report the behavior of $S$ for $\phi\in[0,\pi/2]$). Figure~\ref{fig_2} shows the sloppiness $S$ as a function of $\phi$ for different combinations of the $\tau$ and $\eta$. The analysis performed for both the low-photon ($N=1$) and high-photon ($N=10$). We have compared the results for coherent-state probe ($\beta\approx0$) and in the presence of a high squeezing contribution ($\beta= 0.9$). Analogous numerical results can be obtained for other choices of the involved parameters. As we can see, the sloppiness always exhibits a  global minimum at $\phi=0$: here we have the highest estimation performance, thus we choose $\phi=0$ as the optimal measurement phase and use this value throughout the following analysis.
\begin{figure}[h!]
\centering
\includegraphics[width=.335\textwidth]{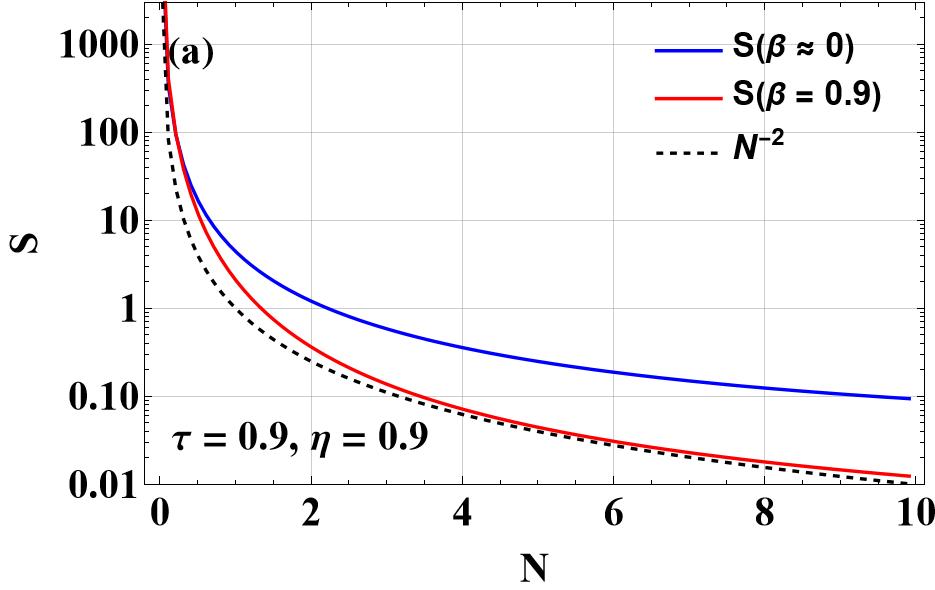}
\includegraphics[width=.315\textwidth]{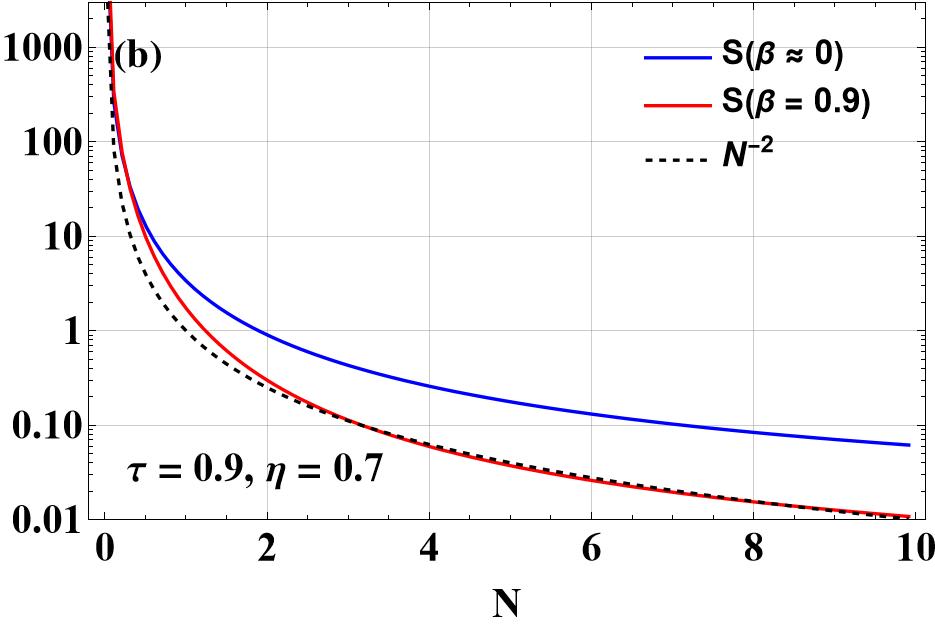}
\includegraphics[width=.315\textwidth]{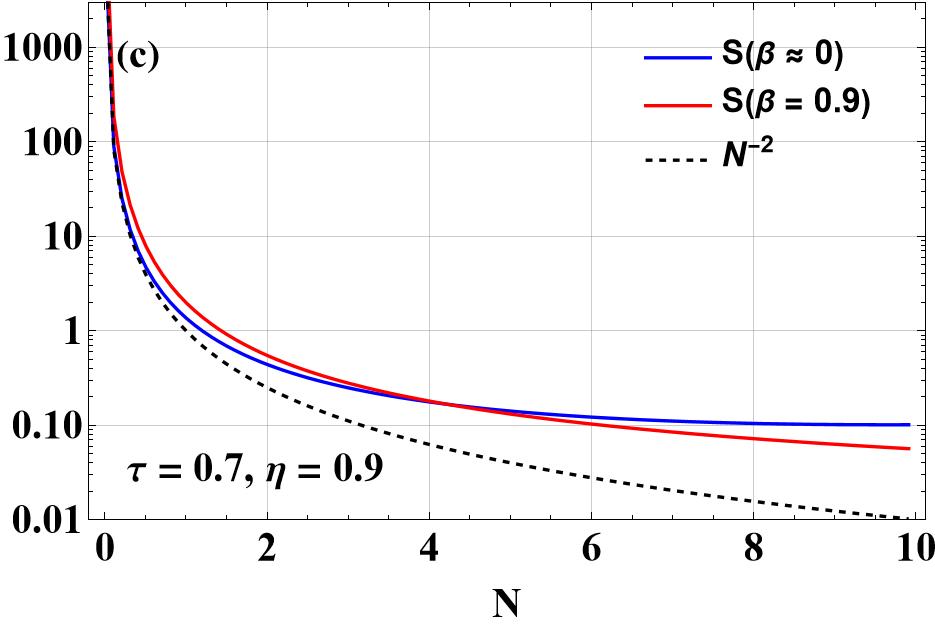}
\caption{\label{fig_3} Sloppiness $S$ as a function of the input energy $N$ for different values of $\tau$ and $\eta$ with nearly no squeezing ($\beta \approx 0 $), and with a relevant squeezing contribution ($\beta = 0.9 $). The dashed lines denote the scaling $N^{-2}$, that we plotted to better understand the numerical results.}
\end{figure} 

Having identified the optimal measurement quadrature, we next examine how the probe-state resources influence parameter's estimation. Figure~\ref{fig_3} shows the sloppiness as a function of the total mean photon number for three different choices of $\tau$ and $\eta$. For $(\tau,\eta)=(0.9,0.9)$ and $(0.9,0.7)$, the sloppiness decreases monotonically with increasing photon number, and the reduction is significantly enhanced by squeezing ($\beta=0.9$). In the panels of Fig.~\ref{fig_3} we also plotted the scaling $N^{-2}$ to better appreciate the numerical results. In contrast, for $(\tau,\eta)=(0.7,0.9)$ the improvement due to squeezing is much less pronounced, indicating that the benefit of nonclassical resources depends on the operating point of the photon-subtraction device.
In this framework, we have performed a systematic numerical analysis, which establishes that the amplitude quadrature is always the optimal measurement basis and that squeezing is most beneficial in the high-transmissivity regime, while in other regimes it may not be equally effective in reducing sloppiness. For these reasons, in the rest of the Section we compare two different estimation strategies, the one based on the joint estimation of the two parameters and the other on a {\textit sequential} estimation strategy in which we first estimate $\tau$ and, the, we use this information to estimate $\eta$.

\subsection{Joint Estimation}
We first consider the joint (simultaneous) estimation strategy, in which both unknown parameters are inferred from the same measurement outcomes. The corresponding Cramér-Rao bound is given by  
\begin{equation}
\Delta^2\tau + \Delta^2\eta \geq \frac{C_F}{M},
\end{equation}
where $C_F=\mathrm{Tr}(\F^{-1})$ and $M$ is the number of measurement. The individual estimation precisions are determined by the diagonal elements of the inverse Fisher information matrix, $\left[\F^{-1}\right]_{\tau\tau}$ and $\left[\F^{-1}\right]_{\eta\eta}$.
\begin{figure}[h]
\centering
\includegraphics[width=.32\textwidth]{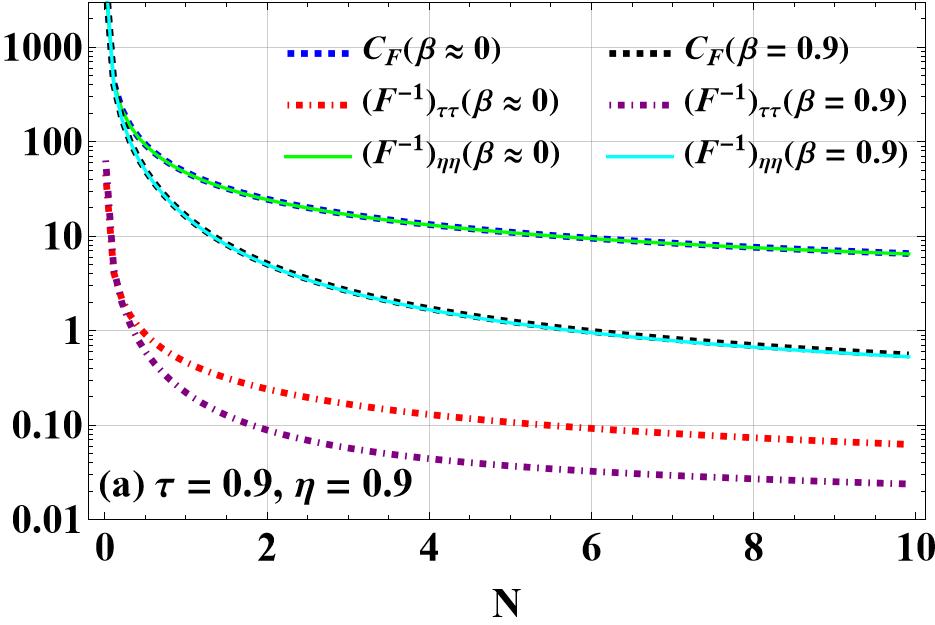}
\includegraphics[width=.32\textwidth]{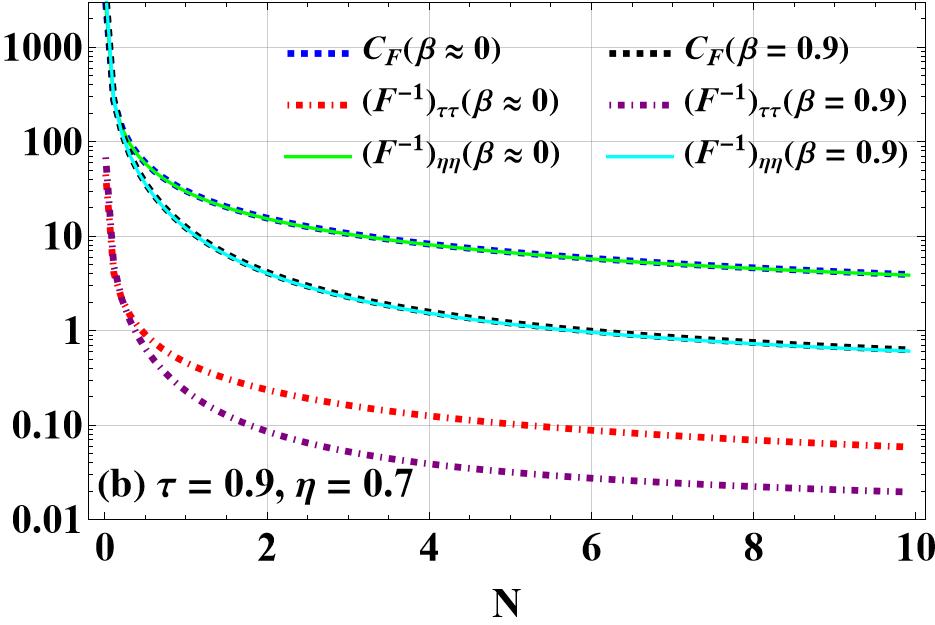}
\includegraphics[width=.32\textwidth]{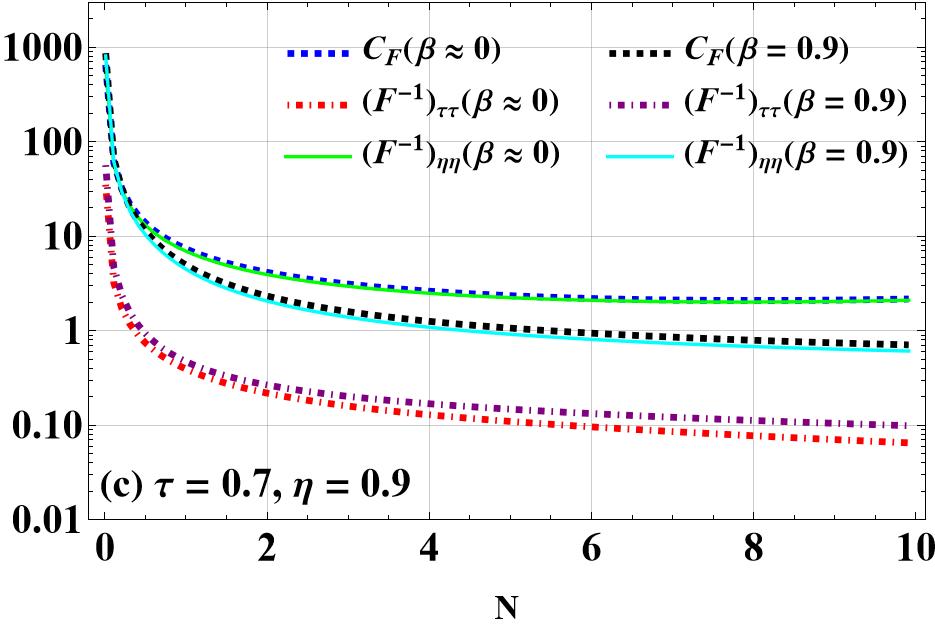}
\caption{\label{fig_4}Plots of $C_F$, $\left[\F^{-1}\right]_{\tau\tau}$, and $\left[\F^{-1}\right]_{\eta\eta}$ as functions of $N$. Plots (a), (b), and (c) correspond to the parameter sets $(\tau, \eta) = (0.9, 0.9)$, $(0.9, 0.7)$, and $(0.7, 0.9)$, respectively.}
\end{figure}

Fig.~\ref{fig_4} shows the dependence of $C_F$, $\left[\F^{-1}\right]_{\tau\tau}$ and $\left[\F^{-1}\right]_{\eta\eta}$ on the total mean photon number for different values of $\tau$ and $\eta$. In all cases, the estimation bounds decrease monotonically with increasing photon number, indicating that higher probe energies improve the simultaneous estimation of both parameters. Moreover, the use of squeezing ($\beta=0.9$) consistently reduces $C_F$ as well as the individual estimation bounds compared with the nearly coherent probe ($\beta\approx0$). Among the parameter sets considered, the smallest values of $C_F$, $\left[\F^{-1}\right]_{\tau\tau}$, and $\left[\F^{-1}\right]_{\eta\eta}$ are obtained for $(\tau,\eta)=(0.7,0.9)$, demonstrating that the joint estimation protocol performs most favorably in this realistic regime.

\subsection{Sequential (Conditional) Estimation}

\begin{figure}[h]
\centering
\includegraphics[width=.33\columnwidth]{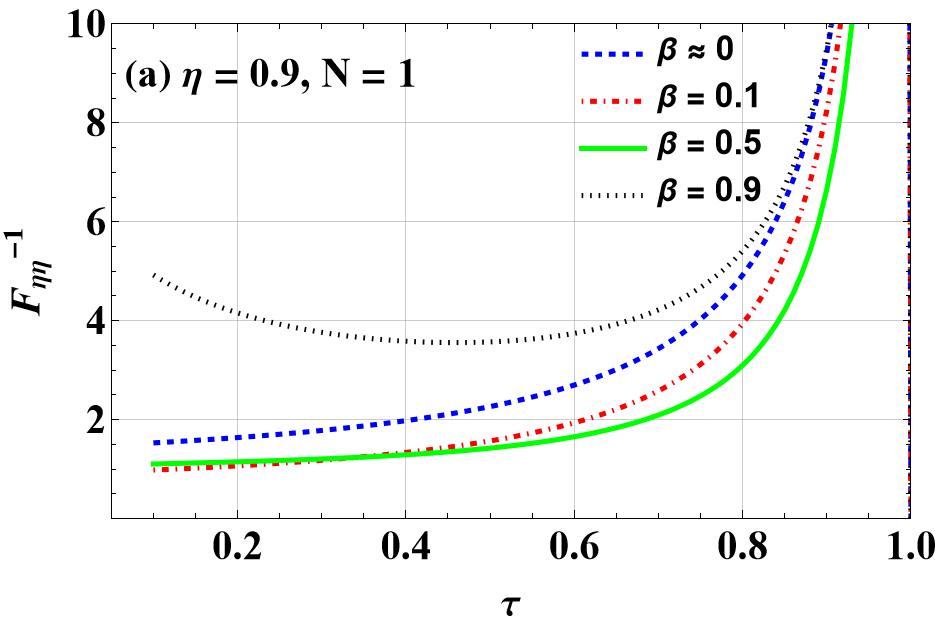}
\includegraphics[width=.33\columnwidth]{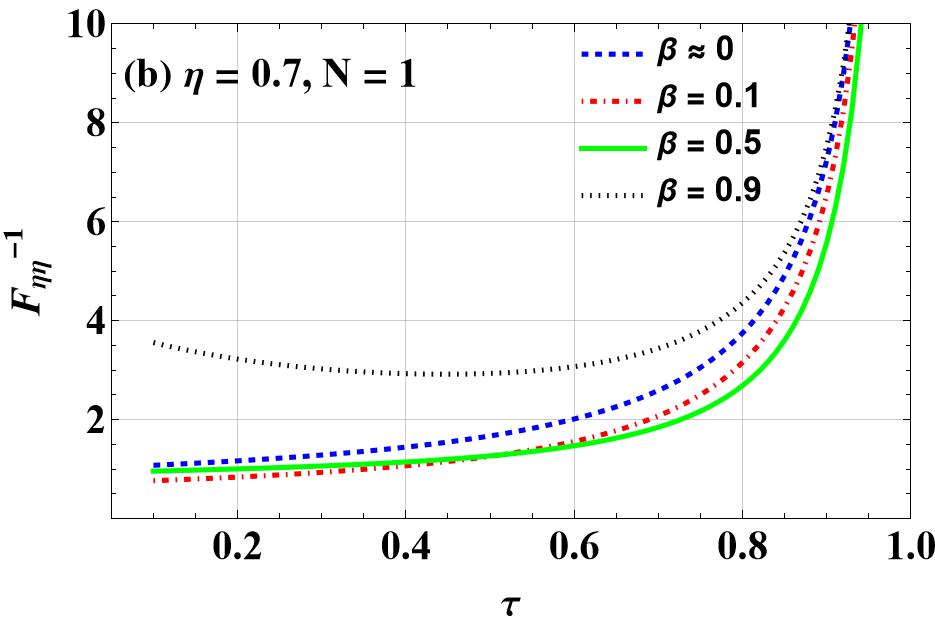}
\includegraphics[width=.33\columnwidth]{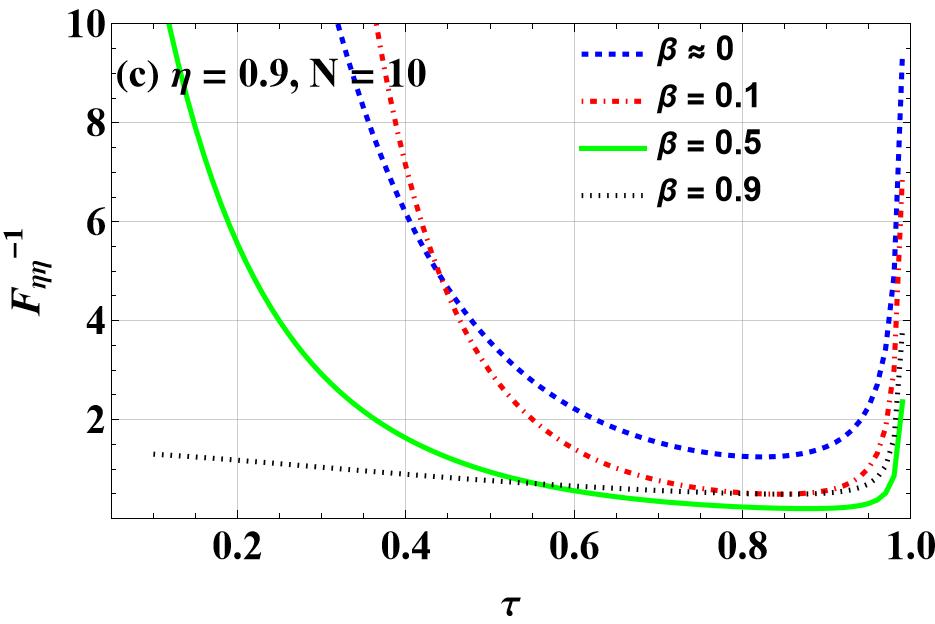}
\includegraphics[width=.33\columnwidth]{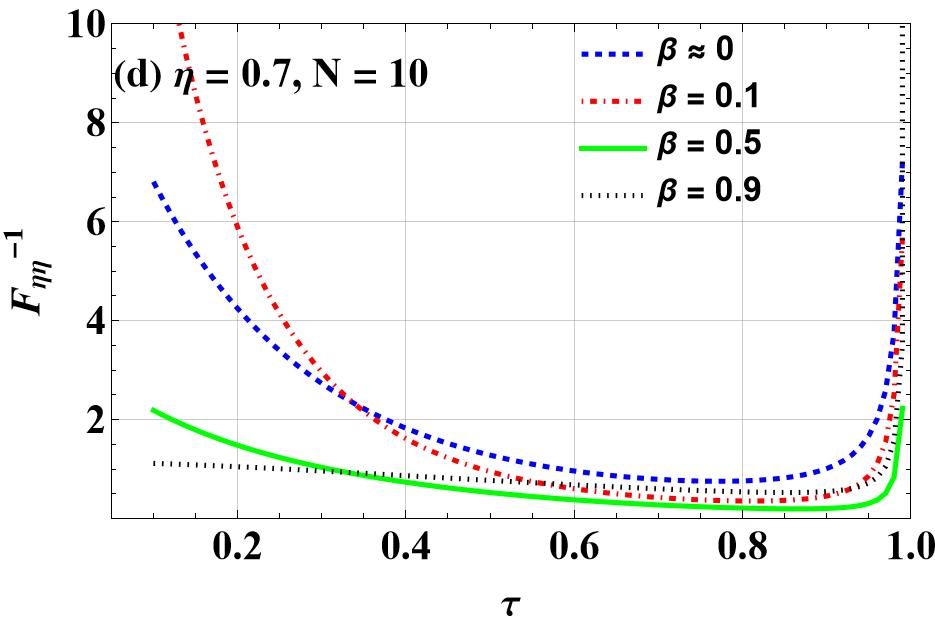}
\caption{\label{fig_5} $({\F}_{\eta\eta})^{-1}$ as a function of the beam-splitter transmissivity $\tau$ for different squeezing fractions $\beta$. Panels (a) and (b) correspond to $N=1$, while panels (c) and (d) correspond to $N=10$. The detector quantum efficiencies are $\eta=0.9$ and $\eta=0.7$ as reported in the corresponding panels.}
\end{figure}
Although joint estimation provides the best overall precision, it is instructive to compare it with a sequential (conditional) estimation strategy. In this approach, the BS transmissivity $\tau$ is estimated first, after which the detector quantum efficiency $\eta$ is estimated by modulating the transmissivity at another, appropriate, operating point $\tau^*$. The corresponding lower bound is given by
\begin{equation}
\Delta^2\tau + \Delta^2\eta \geq \frac{1}{M}
\left\{\left[\F^{-1}\right]_{\tau\tau}+\frac{1}{\left[\F(\tau^*)\right]_{\eta\eta}}\right\},
\end{equation}
where $\left[\F(\tau^*)\right]_{\eta\eta}$ denotes the Fisher information for estimating $\eta$ at the chosen (known) value $\tau=\tau^*$.

To optimize this protocol, we first determine the value of $\tau$ that minimizes the estimation uncertainty of $\eta$. Figure~\ref{fig_5} shows $\left[\F^{-1}\right]_{\eta\eta}$ as a function of $\tau$ for different probe energies and squeezing fractions. For nearly coherent probes ($\beta\approx0$), the optimal transmissivity depends on the photon number, shifting from $\tau\simeq0.1$ in the low-photon regime to $\tau\simeq0.9$ for higher photon numbers. In contrast, for strongly squeezed probes ($\beta=0.9$), the optimum occurs around $\tau\simeq0.5$, indicating that squeezing modifies the optimal operating point for estimating the detector efficiency.

\begin{figure}[h]
\centering
\includegraphics[width=.33\columnwidth]{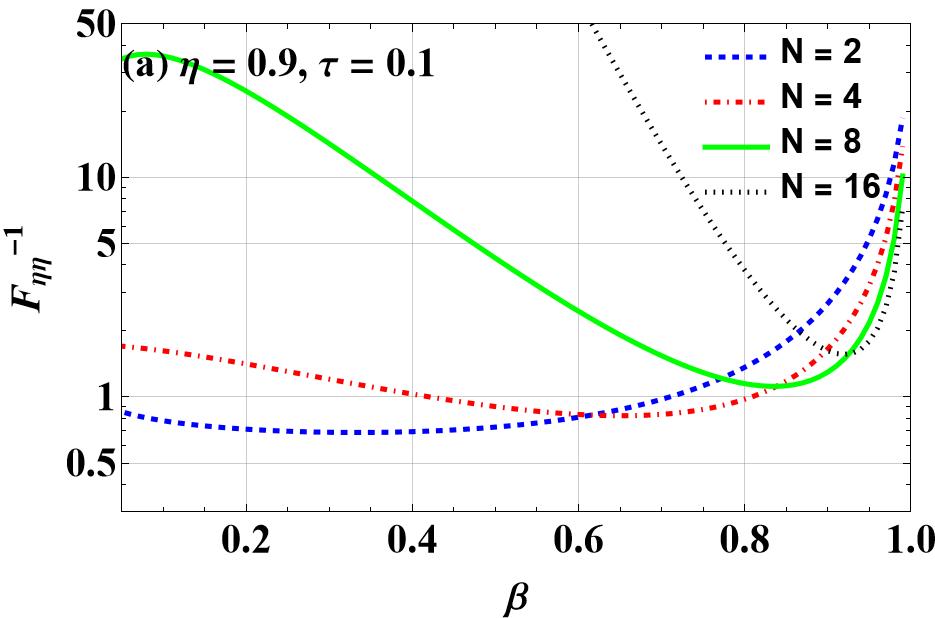}
\includegraphics[width=.33\columnwidth]{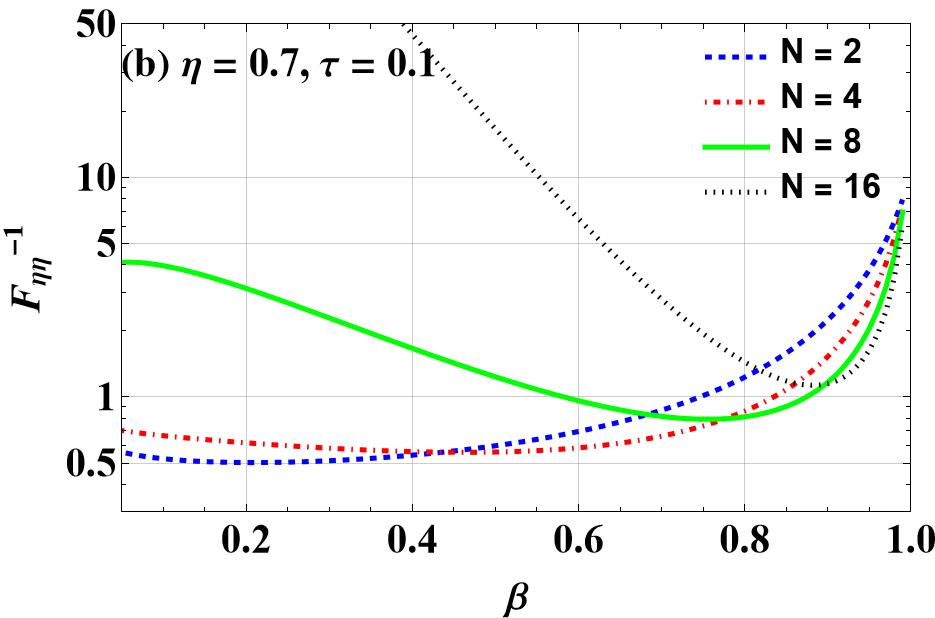}
\includegraphics[width=.33\columnwidth]{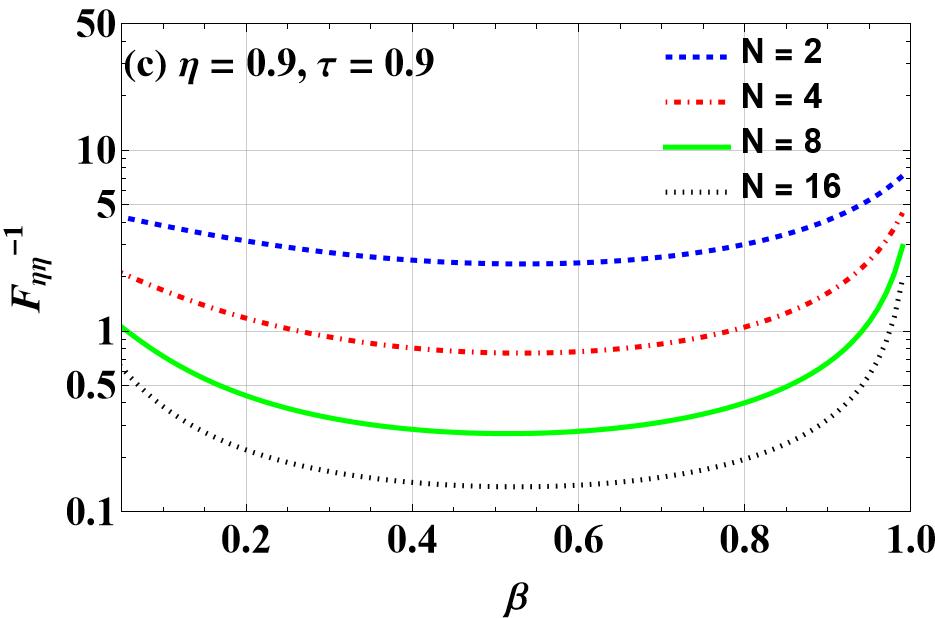}
\includegraphics[width=.33\columnwidth]{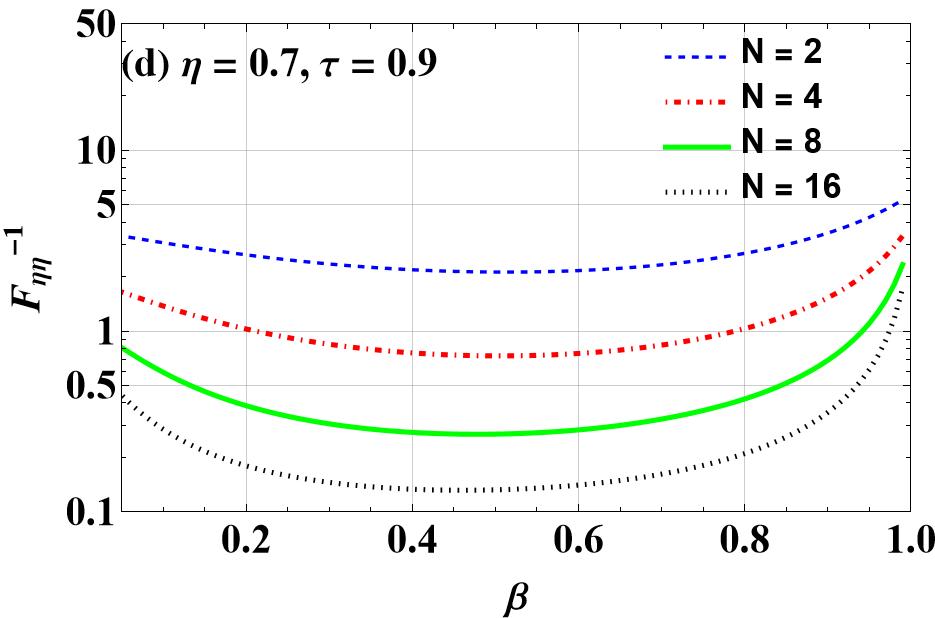}
\caption{\label{fig_6} $\left[{\F}\right]_{\eta\eta}^{-1}$ as a function of the squeezing fraction $\beta$ for different mean photon numbers. Panels (a) and (c) are for $\eta=0.9$, whereas panels (b) and (d) are for $\eta=0.7$. Panels (a) and (b) correspond to the optimized transmissivity $\tau=0.1$, while panels (c) and (d) correspond to $\tau=0.9$.}
\end{figure}
After determining the optimal operating point, we next investigate whether the squeezing fraction can also be optimized. Figure~\ref{fig_6} reports $\left[\F^{-1}\right]_{\eta\eta}$ as a function of $\beta$ for the optimized values of $\tau$. For $\tau=0.1$, $\left[\F^{-1}\right]_{\eta\eta}$ varies monotonically with $\beta$, and no distinct optimum is observed over the range considered. In contrast, for $\tau=0.9$, the estimation bound exhibits a clear minimum around $\beta\simeq0.5$ for both values of the detector efficiency. Thus, in the high-transmissivity regime, a balanced combination of coherent and squeezed photons provides the highest precision for estimating $\eta$.

Having optimized the sequential protocol with respect to both $\tau$ and $\beta$, we now compare its performance directly with the joint estimation strategy.

\subsection{Comparison between Joint and Sequential Estimation}
\begin{figure}[h]
\centering
\includegraphics[width=.32\textwidth]{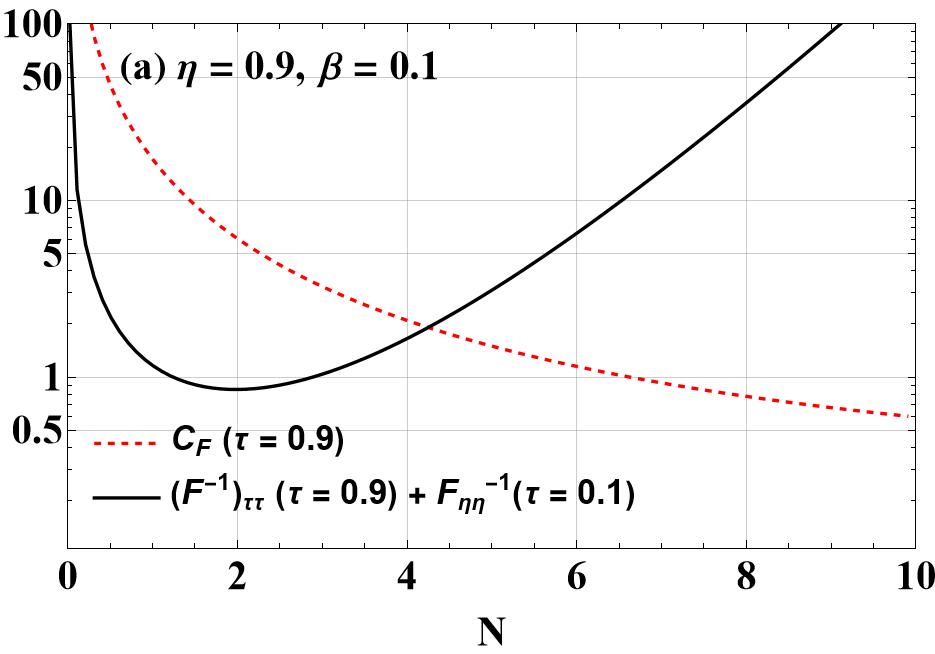}
\includegraphics[width=.32\textwidth]{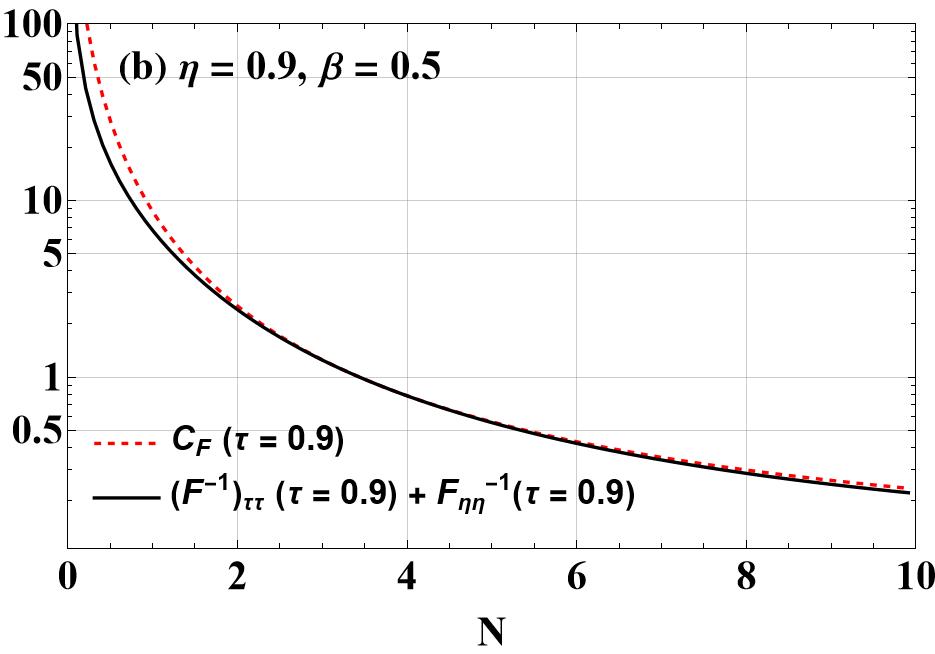}
\includegraphics[width=.32\textwidth]{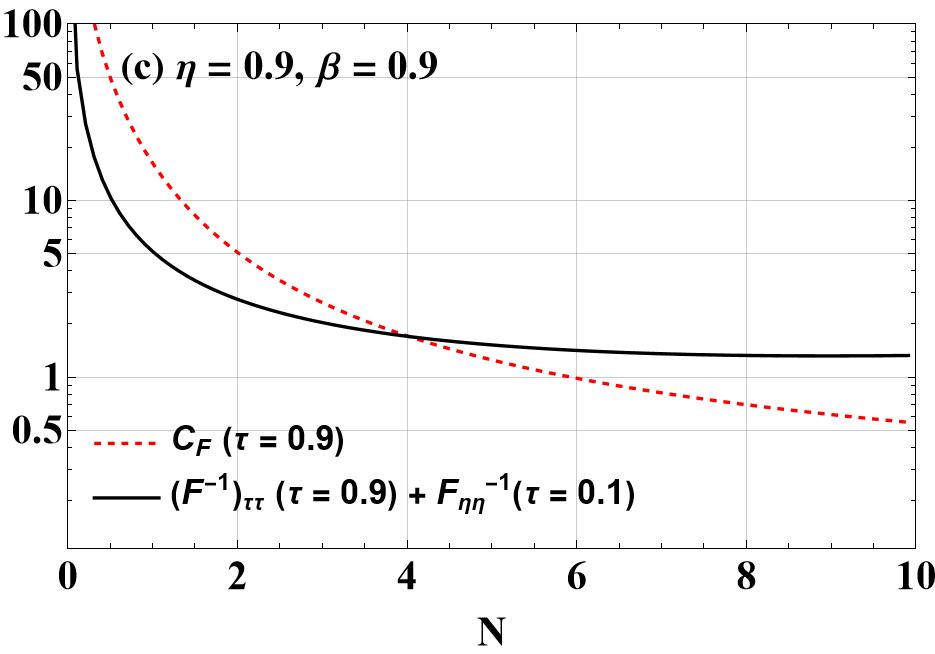}
\caption{\label{fig_7} Comparison of the lower bounds for joint estimation, $C_F$, and sequential estimation,
$\left[\F^{-1}\right]_{\tau\tau}+\left[\F(\tau^*)\right]_{\eta\eta}^{-1}$, as functions of the mean photon number.
The sequential protocol is evaluated at the optimized values of $\tau$ and $\beta$. Panels
(a) and (c) correspond to $\tau^*=0.1$, while panel (b) corresponds to $\tau^*=0.9$}.
\end{figure}

Having optimized the sequential protocol over both $\tau$ and $\beta$, we now compare it with the joint estimation strategy. Figure~\ref{fig_7} shows the corresponding lower bounds as functions of the mean photon number.

 For all parameter sets considered, the joint-estimation bound, $C_F$, remains lower than the sequential bound, $\left[\F^{-1}\right]_{\tau\tau}+\left[\F(\tau^*)\right]_{\eta\eta}^{-1}$, demonstrating that estimating both parameters simultaneously provides higher precision than estimating them sequentially. The advantage persists over the entire range of probe energies and becomes more pronounced as the photon number increases. These results establish joint estimation as the more efficient strategy for the simultaneous characterization of the beam-splitter transmissivity and detector quantum efficiency.

\section{Conclusion}\label{Conclusion}
In this work, we have proposed a practical in situ protocol for the simultaneous, namely, joint estimation of the BS transmissivity and the quantum efficiency of the heralding detector in a photon-subtraction device. The proposed scheme relies exclusively on experimentally accessible data, namely the click statistics of the on/off detector and homodyne measurements performed on the transmitted output mode, thereby eliminating the need for independent classical calibration procedures.

Within the framework of classical multi-parameter estimation theory, we derived the total Fisher information matrix by combining the information obtained from the detector click statistics with that extracted from the conditional homodyne measurements \textcolor{black}{when the input probe state is a displaced squeezed state with real squeezing parameter}. We have employed the sloppiness parameter as a quantitative measure of parameter distinguishability and used it to analyze the degree of correlation between the two unknown parameters. Our results show that the sloppiness attains its minimum at the homodyne phase $\phi = 0$, indicating that this measurement configuration provides the optimal operating point for the simultaneous estimation protocol.

We have then investigated the influence of the probe photon number, the squeezing fraction, the BS transmissivity, and the detector efficiency on the estimation precision. The analysis demonstrates that both the probe-state resources and the operating parameters of the photon-subtraction device significantly affect the achievable estimation bounds. In particular, the actual value of the BS transmissivity and the optimization of the squeezing fraction identifies the operating regimes that provide the highest estimation precision for the considered estimation strategies.

Finally, we have compared joint estimation with sequential estimation using the optimized operating parameters. The results consistently show that the joint estimation strategy yields a lower estimation bound than the sequential approach over the investigated parameter range, demonstrating that simultaneous estimation provides a more efficient characterization of the photon-subtraction device.

The estimation protocol presented here is based entirely on measurements that are available in quantum optical experiments and can therefore be implemented without modifying the standard photon-subtraction setup. We expect that the proposed approach will provide a useful and experimentally feasible method for the characterization and calibration of photon-subtraction devices employed in continuous-variable quantum information processing, quantum state engineering, and quantum metrology.

\begin{acknowledgements}
PS thanks the no-profit organization {\em Comitato Quantum}, which supported this work through a travel grant.
\end{acknowledgements}

\appendix

\section{Gaussian Evolution and Conditional State}
\label{AppA}
In this Appendix, we derive the conditional Gaussian state used in
Sec.~\ref{Sec:Evolution}. The output covariance matrix and displacement
vector after the beam splitter are obtained from \cite{olirev2012}
\begin{align}
\gamma_{\rm out}
&=
S_{\rm BS}
\begin{pmatrix}
\gamma_1&0\\
0&\gamma_2
\end{pmatrix}
S_{\rm BS}^{\intercal}
= 
\begin{pmatrix}
\gamma_3&\gamma_{34}\\
\gamma_{34}&\gamma_4
\end{pmatrix}
\label{A1}
\\[1ex]
d_{\rm out}
&=
S_{\rm BS}
\begin{pmatrix}
d_1\\
d_2
\end{pmatrix} = \begin{pmatrix}
d_3\\
d_4
\end{pmatrix},
\label{A2}
\end{align}
where $S_{\BS}$ is the BS symplectic transformation  
\begin{equation}
S_{\BS}=
\begin{pmatrix}
\sqrt{\tau}\,\Id_2 & \sqrt{1-\tau}\,\Id_2 \\[1ex]
-\sqrt{1-\tau}\,\Id_2 & \sqrt{\tau}\,\Id_2
\end{pmatrix},\label{BS}
\end{equation}
being $\tau$ the BS transmissivity.
From Eq.~(\ref{A1}), the covariance matrices of the transmitted mode,
reflected mode, and their correlation matrix are
\begin{align}
\gamma_3
&=
\frac12
\begin{pmatrix}
\tau e^{-2\lambda}+1-\tau&0\\
0&\tau e^{2\lambda}+1-\tau
\end{pmatrix},
\\[1ex]
\gamma_4
&=
\frac12
\begin{pmatrix}
(1-\tau)e^{-2\lambda}+\tau&0\\
0&(1-\tau)e^{2\lambda}+\tau
\end{pmatrix},
\\[1ex]
\gamma_{34}
&=
-\frac{\sqrt{\tau(1-\tau)}}2
\begin{pmatrix}
e^{-2\lambda}-1&0\\
0&e^{2\lambda}-1
\end{pmatrix}.
\label{A3}
\end{align}
also we have
\begin{align}
d_3 = \sqrt{\tau}\begin{pmatrix} x_0 \\ y_0 \end{pmatrix}\, \quad\mbox{and}\quad
d_4 = -\sqrt{1-\tau}\begin{pmatrix} x_0 \\ y_0 \end{pmatrix}.
\end{align}
The Gaussian positive operator-valued measure describing the ``off'' event has the following covariance matrix and displacement vector
\begin{align}
\gamma_{P_0}
= 
\frac{2-\eta}{2\eta}\,\Id_2,\quad\mbox{and}\quad
d_{P_{0}} = \begin{pmatrix}
0\\
0
\end{pmatrix},
\label{A4}
\end{align}
where $\eta$ is the detector quantum efficiency.
Using the Gaussian conditioning formula \cite{olirev2012}
\begin{align}
\gamma_0
&=
\gamma_3
-
\gamma_{34}
(\gamma_4+\gamma_{P_0})^{-1}
\gamma_{34}^{\intercal},
\label{A5}
\\[1ex]
d_0
&=
d_3
-
\gamma_{34}
(\gamma_4+\gamma_{P_0})^{-1}
(d_4-d_{P_0}),
\label{A6}
\end{align}
one straightforwardly obtains Eqs.~(\ref{condcov}) and
(\ref{conddisp}) of the main text.

\section{Derivation of the Fisher Information Matrix}
\label{AppB}
This Appendix summarizes the derivation of the Fisher information matrix used in Section \ref{Methodology}. Since the estimation protocol combines homodyne measurements with the heralding detector statistics, we first derive the Fisher information associated with the conditional homodyne distributions and then combine it with the information carried by the detector outcomes.
\subsection{Fisher Information from the ``OFF'' Events}\label{AppC1}

For the ``off'' events, the homodyne distribution is Gaussian. The probability distribution for measuring value $x$ in quadrature $q_\phi$ is
\begin{equation}
P(x|\boldsymbol{\theta}) = \frac{1}{\sqrt{2\pi\sigma_\phi^2}} \exp\left[ -\frac{(x - \mu_\phi)^2}{2\sigma_\phi^2} \right]
\end{equation}
where
$\mu_\phi$ and $\sigma_\phi^2$ are given in Eq.~(\ref{MeanQuadrature}) and Eq.(\ref{VarianceQuadrature}) respectively.  
The Fisher information is therefore
\begin{equation}
\left[\Info^{{\rm hd(off)}}\right]_{jk}
=
\frac{\partial_j\mu_\phi\, \partial_k\mu_\phi}{\sigma_\phi^2}
+
\frac{\partial_j\sigma_\phi^2\,\partial_k\sigma_\phi^2}{2\sigma_\phi^4},
\label{B6OFF}
\end{equation}
where the derivatives are evaluated using Eqs.~(\ref{MeanQuadrature}) and (\ref{VarianceQuadrature}). The resulting matrix elements are
\begin{align}
\Info_{\tau\tau}
&=
\left(\frac{\partial_\tau\mu_\phi}{\sigma_\phi}\right)^2
+
\frac12 \left(\frac{\partial_\tau\sigma_\phi^2}{\sigma_\phi^2}\right)^2
\\
\Info_{\tau\eta}
&= \Info_{\eta\tau} =
\frac{\partial_\tau\mu_\phi\, \partial_\eta\mu_\phi}{\sigma_\phi^2}
+
\frac{\partial_\tau\sigma_\phi^2\, \partial_\eta \sigma_\phi^2}{2\sigma_\phi^4},
\\
\Info_{\eta\eta}
&=
\left(\frac{\partial_\eta\mu_\phi}{\sigma_\phi}\right)^2
+
\frac12 \left(\frac{\partial_\eta\sigma_\phi^2}{\sigma_\phi^2}\right)^2.
\end{align}
The required derivatives are obtained directly from
Eqs.~(\ref{MeanQuadrature}) and (\ref{VarianceQuadrature}).

\subsection{Total Fisher Information}\label{AppC2}

The total Fisher information matrix ${\mathcal F}$ follows from the joint probability distributions, namely
\begin{equation}
P(\chi,x|\boldsymbol{\theta})
=
P(\chi|\boldsymbol{\theta})\,
p_\chi(x|\boldsymbol{\theta}),\quad 
\label{B1}
\end{equation}
with $\chi=0,1$ (``off'', ``on'') and $x\in {\mathbbm R}$, whose logarithmic derivative are separated into detector and homodyne contributions
\begin{equation}
\partial_j
\log P(\chi,x|\boldsymbol{\theta})
=
\partial_j
\log P(\chi|\boldsymbol{\theta})
+
\partial_j
\log p_\chi(x|\boldsymbol{\theta}).
\label{B2}
\end{equation}

Using the definition of the Fisher information matrix,
\begin{equation}
\left[{\mathcal F}\right]_{jk}
=
\mathbb{E}
\!\left[
\partial_j
\log P(\chi,x|\boldsymbol{\theta})
\,
\partial_k
\log P(\chi,x|\boldsymbol{\theta})
\right],
\end{equation}
and substituting Eq.~(\ref{B2}), we obtain
\begin{align}
\left[{\mathcal F}\right]_{jk}
&=
\mathbb{E}\!\left[
\partial_j\log P(\chi|\boldsymbol{\theta})
\,\partial_k\log P(\chi|\boldsymbol{\theta})
\right]
\nonumber\\[1ex]
&\quad\quad+
\mathbb{E}\!\left[
\partial_j\log P(\chi|\boldsymbol{\theta})
\,\partial_k\log p_\chi(x|\boldsymbol{\theta})
\right]
\nonumber\\[1ex]
&\quad\quad\quad+
\mathbb{E}\!\left[
\partial_j\log p_\chi(x|\boldsymbol{\theta})
\,\partial_k\log P(\chi|\boldsymbol{\theta})
\right]
\nonumber\\[1ex]
&\quad\quad\quad\quad+
\mathbb{E}\!\left[
\partial_j\log p_\chi(x|\boldsymbol{\theta})
\,\partial_k\log p_\chi(x|\boldsymbol{\theta})
\right].
\label{B3}
\end{align}

The second and third terms correspond to the correlations between the score functions associated with the on/off detection and the conditional homodyne measurement. These mixed contributions vanish because the conditional score function has zero mean. Therefore,
\begin{equation}
\mathbb{E}_{x|\chi}
\!\left[
\partial_j
\log p_\chi(x|\boldsymbol{\theta})
\right]
=
0,
\label{B4}
\end{equation}
which implies
\begin{equation}
\mathbb{E}
\!\left[
\partial_j\log P(\chi|\boldsymbol{\theta})
\,
\partial_k\log p_\chi(x|\boldsymbol{\theta})
\right]
=
0,
\label{B5}
\end{equation}
and similarly for the remaining mixed term. Consequently, the total Fisher information is the sum of the Fisher information associated with the on/off statistics and the conditional homodyne measurement, leading to Eq.~(\ref{TotalFisher}), where
\begin{align}
\left[\mathcal{H}^{\text{on/off}}(\boldsymbol{\theta})\right]_{jk}
&=
\mathbb{E}\big[
\partial_j \log P(\chi|\boldsymbol{\theta})\,
\partial_k \log P(\chi|\boldsymbol{\theta})
\big],\\[1ex]
\left[\Info^{{\rm hd}(\chi)}(\boldsymbol{\theta})\right]_{jk}
&=
\mathbb{E}_{x|\chi}\big[
\partial_j\log p_\chi(x|\boldsymbol{\theta})\,
\partial_k \log p_\chi(x|\boldsymbol{\theta})
\big].
\end{align}

For the Gaussian conditional state, the conditional covariance matrix is obtained using the Schur-complement method \cite{Zhang2005}, as given in Eq.~(\ref{A5}). The corresponding on/off detection probability is
\begin{equation}
P(0|\boldsymbol{\theta})
=
\frac{2\exp
\left[
-\frac12
(d_4-d_{P_0})^\intercal
(\gamma_4+\gamma_{P_0})^{-1}
(d_4-d_{P_0})
\right]
}
{\sqrt{\det(\gamma_4+\gamma_{P_0})}},
\label{Poff}
\end{equation}
with $P(1|\boldsymbol{\theta}) = 1 - P(0|\boldsymbol{\theta})$.

\section{Behavior of the Sloppiness Parameter at Higher Photon Numbers ($N >10$)}
In the main text, the analysis was restricted to experimentally relevant probe energies $N\leq10$. To investigate whether the observed trends persist for larger probe energies, we extend the analysis to the higher-photon-number regime $N >10$.

\begin{figure}[h]
\centering
\includegraphics[width=.33\columnwidth]{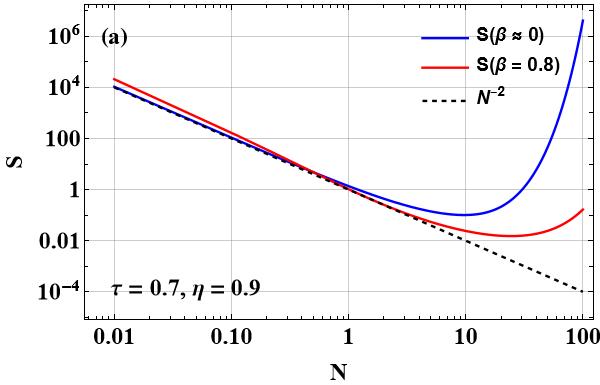}
\includegraphics[width=.33\columnwidth]{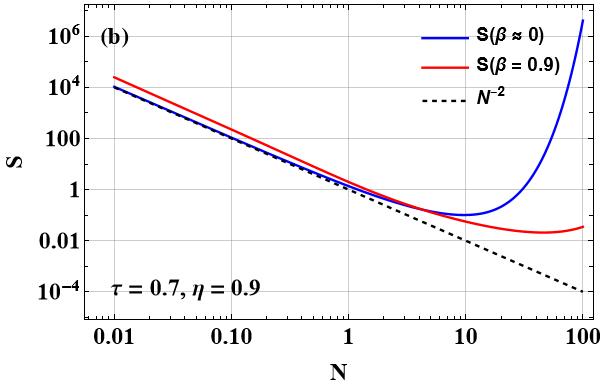}
\caption{\label{fig_8}
Sloppiness  $S$ with the mean photon number for coherent ($\beta=0$) and squeezed ($\beta=0.8$ and $0.9$) probes at $(\tau,\eta)=(0.7,0.9)$. The figure illustrates the behavior of $S$ in the very high-photon-number regime. }
\end{figure}

Figure~\ref{fig_8} further illustrates the dependence of the sloppiness parameter on the probe energy for coherent ($\beta=0$) and squeezed ($\beta=0.8$ and $0.9$) probes at the representative operating point $(\tau,\eta)=(0.7,0.9)$. Consistent with the results presented in the main text, the sloppiness initially decreases with increasing photon number, indicating improved distinguishability between the estimated parameters. However, beyond $N>10$, the sloppiness starts to increase, suggesting that the improvement in estimation precision saturates at higher photon numbers. Throughout the range considered, squeezed probes maintain lower values of $S$ than the coherent probe, confirming the advantage of squeezing in reducing parameter correlations.

\bibliography{reference}
\end{document}